\documentclass[preprint,12pt]{elsarticle}
\usepackage[utf8]{inputenc}

\usepackage{lineno}

\usepackage{pgfplots}
\usetikzlibrary{positioning}
\tikzstyle{process} = [rectangle, minimum width=3cm, minimum height=1cm, text centered, text width=5cm, draw=black, fill=orange!0]
\tikzstyle{arrow} = [thick,->,>=stealth]
\usepackage{graphicx}
\graphicspath{{figures/}}

\usepackage{tabularx}
\usepackage{longtable}
\usepackage{booktabs}

\usepackage{url}
\usepackage{hyperref}

\begin{document}
\begin{frontmatter}
\title{The Use of Public Data and Free Tools in National CSIRTs' Operational Practices: A Systematic Literature Review}

\author[kent,myCERT]{Sharifah Roziah Binti Mohd Kassim\corref{cor}}
\ead{sm2212@kent.ac.uk}
\author[kent]{Shujun Li}
\ead{S.J.Li@kent.ac.uk}
\author[kent]{Budi Arief}
\ead{B.Arief@kent.ac.uk}
\address[kent]{Institute of Cyber Security for Society (iCSS) \& School of Computing, University of Kent, UK}
\address[myCERT]{Malaysia Computer Emergency Response Team (MyCERT), CyberSecurity Malaysia, Malaysia}

\begin{abstract}
Computer Security Incident Response Teams (CSIRTs) have been established at national and organisational levels to coordinate responses to computer security incidents. It is known that many CSIRTs, including national CSIRTs, routinely use public data, including open-source intelligence (OSINT) and free tools, which include open-source tools in their work. However, we observed a lack of public information and systematic discussions regarding how national CSIRTs use and perceive public data and free tools in their operational practices. Therefore, this paper provides a systematic literature review (SLR) to comprehensively understand how national CSIRTs use and perceive public data and free tools in facilitating incident responses in operations. Our SLR method followed a three-stage approach: 1) a systematic search to identify relevant publications from websites of pertinent CSIRT organisations, 2) a conventional SLR into the research literature, and 3) synthesise data from stages one and two to answer the research questions. In the first stage, we searched the websites of 100 national CSIRTs and 11 cross-CSIRT organisations to identify relevant information about national CSIRTs. In the second stage, we searched a scientific database (Scopus) to identify relevant research papers. Our primary finding from the SLR is that most discussions concerning public data and free tools by national CSIRTs are incomplete, ad hoc, or fragmented. We also discovered a lack of discussions on how the staff of national CSIRTs perceive the usefulness of public data and free tools to facilitate incident responses. Such gaps can prevent us from understanding how national CSIRTs can benefit from public data and free tools and how other organisations, individuals and researchers can help by providing such data and tools to improve national CSIRTs' operation. These findings call for more empirical research on how national CSIRTs use and perceive public data and free tools and how such data and tools can be leveraged to improve the overall incident responses at national CSIRTs and other incident response organisations.
\end{abstract}


\end{frontmatter}

\section{Introduction}
\label{sec:Introduction-SLR}

Cyber incidents continue to increase globally, exemplified by the increase in ransomware incidents with an almost 13\% increase compared to the year 2020~\cite{Verizon2022DBIR}. Furthermore, increased threat actors of all motivation and skill levels have contributed to increased cyber-attacks across organisations, sectors and geographies throughout 2021~\cite{Pwc2021}. The Covid-19 pandemic has somehow increased the exposure of risks, vulnerabilities, and cyber threats due to increased remote working activities~\cite{APCERT2021} and other online transactions that led to unforeseen new vulnerabilities~\cite{korn2021jack}. This causes a further increase in the number and severity of cyber attacks worldwide~\cite{iakovakis2021analysis}. By looking at the current threat landscape, organisations need to be more ``cyber-resilient'' with more effective approaches to defending against cyber attacks~\cite{Tariq2013}. Although the system and network defence are essential, responding to attacks in a timely and efficient manner -- quite often, facilitated by a dedicated incident response team~\cite{huck2022wake} is equally important~\cite{costa2021practical}. This includes having effective incident response teams with experience, technical skills, and capacity to respond to cyber incidents~\cite{EuropeanCommission2009} to help lower risks to organisations~\cite{ruefle2014computer}. The growth of cyber security science, which is trying to establish procedures, mechanisms, and techniques to face cyber attacks, shows the relevance of cyber security incident response teams (CSIRT) in responding to cyber attacks~\cite{espin2020guidelines}. 

An incident response team is vital for mitigating cyber incidents and minimising the damage, cost and recovery time. Such a team is also instrumental in finding and fixing the root cause(s) of incidents and subsequently preventing future attacks~\cite{suhail2022security}. Such a team is also called a \textit{computer emergency response (or readiness) team (CERT)}~\cite{ahmad2011organisation}, a \textit{computer security incident response (or readiness) team (CSIRT)}~\cite{ruefle2014computer}, or a \textit{cyber (or computer) incident response (or readiness) team (CIRT)}~\cite{ITU-CIRT}. These acronyms are often used interchangeably in the research literature and by cyber security professionals to refer to cyber security incident response teams. For this article, we use ``CSIRT'' because this is widely used in the research literature and by cyber security professionals.

Many early CSIRTs were established by academic communities, for example, the CERT/CC established at Carnegie Mellon University in the US~\cite{duijnhoven2021stimulating} and the Australian Computer Emergency Response Team -- AusCERT, established at the Queensland University in Australia. Over time, more CSIRTs were set up worldwide, in different sectors and organisations, and at the national level. Some CSIRTs provide full services related to cyber incident management, including reactive services such as incident response with analysis and intrusion detection and proactive services such as vulnerability management, risk assessments, security consulting, and penetration testing~\cite{ortiz2021proposal}. In contrast, many others focus on reactive services only. Organisational or internal CSIRTs (sometimes referred to  as ``enterprise'' CSIRTs) operate at the organisational level -- such as within a private company, a hospital, an enterprise or a government agency~\cite{duijnhoven2021stimulating}. In contrast, national CSIRTs\footnote{For the sake of brevity, in this paper, we use the term ``national'' to refer to both national and regional CSIRTs.} operate at the national level and act as a national point of contact~\cite{ahmad2011organisation} for mitigation of cyber attacks within a specific constituency~\cite{madnick2009}. National CSIRTs also facilitate collaboration and information sharing across nations and organisations~\cite{ahmad2011organisation, ENISA2010, Cheang2009, Roziah2019, Hashem2019, Jalal2017}, including coordination of cyber incident response at the national levels~\cite{IGF2014CSIRT}. Due to the key roles of national CSIRTs, more nations worldwide are encouraged to establish national CSIRTs~\cite{Morgus2015}.

National CSIRTs typically have access to various closed-source data from collaborative partners, including self-reported incident reports by individuals and organisations and cyber intelligence shared by other collaborative CSIRTs~\cite{kassim2014automating, kassim2016exploitation}.  Nonetheless, we have observed a lack of systematic and open discussions on how national CSIRTs use public data and free tools to facilitate cyber incident responses in their operation. This is also reflected in the personal observation of this paper's first author while working at a national CSIRT, whereby this personal observation inspired the study reported in this paper.

This paper uses the term ``public data'' to refer to data available to the public on the Internet for free. The term ``closed-source data'' refers to private or confidential data available to restricted data consumers (e.g., CSIRTs, public bodies, or people with a special security clearance status). The term ``free tools'' is used to refer to free software, open-source tools and ``free online services'' that can be accessed by \emph{anyone for free}\footnote{Some CSIRTs have developed tools that are available to authorised parties (e.g., to other CSIRTs) only. We do not consider such tools as free, regardless of whether they are provided for free to trusted parties.} on the Internet.

This Systematic Literature Review aims \textit{to gain a comprehensive understanding of the use of public data and free tools in the operational practices of national CSIRTs} to \textit{identify areas for future research that help national CSIRTs to improve their capabilities in incident responses}. Two specific research questions (RQs) are specified for the SLR:

\begin{itemize}
\item RQ1: How has past research studied the current practices in national CSIRTs concerning public data and free tools to facilitate incident response?

\item RQ2: How has past research studied national CSIRT staff's perception of the usefulness of public data and free tools in their daily practice?
\end{itemize}

The main contributions of the SLR can be summarised as follows:
\begin{itemize}

\item A general understanding of the operational practices of national CSIRTs and CSIRTs concerning tools and data through a systematic search into research literature and the website of national CSIRTs and cross-CSIRT organisations.

\item An innovative SLR approach -- the ``3-stage'' approach to answering the study's research questions, instead of a traditional SLR. The SLR looked into the websites of relevant organisations (national CSIRTs and cross-CSIRT organisations) and relevant research papers.



\item The study identified research gaps for guiding future research by researchers, stakeholders and other individuals concerning the use of public data and free tools in national CSIRTs' operations.

\item A comprehensive list of reliable and valid free tools, identified through a systematic search into national CSIRTs' websites, Cross-CSIRT organisations and research publications. This list will be published as an open resource for national CSIRTs, CSIRTs and security practitioners for real-world operations to support incident responses.  
\end{itemize} 

The rest of this paper is organised as follows. Section~\ref{sec:Methodology-SLR} describes the methodology adopted for the SLR. The results are described in Section~\ref{sec:Results-SLR}, followed by further discussions on key findings, research gaps, and recommendations in Section~\ref{sec:Discussions-SLR}.

\section{Methodology}
\label{sec:Methodology-SLR}

A systematic literature review (SLR) approach was preferred rather than a non-systematic Literature Review (LR) because the former is a more rigorous and reproducible methodology for conducting literature reviews~\cite{Kitchenham2007}. The PRISMA (Preferred Reporting Items for Systematic Reviews and Meta-Analysis) procedure~\cite{Liberati2009, Kitchenham2007, Higgins2020}, a preferred SLR procedure used in many research fields, was followed in the SLR. The PRISMA procedure consists of four main phases: 1) identification of data sources and candidate data items, 2) screening of candidate data items according to exclusion criteria (based on checking metadata), 3) checking the eligibility of candidate data items according to inclusion criteria (based on reading full-text), and 4) analysis of final included data items.

Traditionally, data items covered in SLRs are research papers identified through systematic searches into scientific databases. To study the RQs in this study, it is considered insufficient to consider just research papers because operational practices of CSIRTs, in general, are often published in other forms of non-research publications (e.g., operational documents, guidelines, articles on websites, training manuals, and even informal online discussions)~\cite{spring2021review}. Furthermore, CSIRT practices cover professional and business aspects; therefore, sources on CSIRTs are not limited to research literature only as very little is available in the research literature~\cite{spring2021review}. 

The SLR was conducted following a three-staged approach:
\begin{itemize} 
\item Stage 1: a systematic search to identify relevant information (web pages and online documents such as technical reports, training manuals, articles, best practices and guidelines) accessible on websites of national CSIRTs and cross-CSIRT organisations.

\item Stage 2: a systematic search to identify relevant research papers in a scientific database.

\item Stage 3:  a synthesis of data items identified from the above two stages to answer the RQs.
\end{itemize}

Sections~\ref{subsec:Stage1} and \ref{subsec:Stage2} explain the detailed processes of the first two stages, respectively, and Section~\ref{sec:Results-SLR} reports the results of our analysis of the data items.

The scope of free tools was limited to those designed for and used by national CSIRTs for incident response activities in their operation. Hence, excluded the following major categories of tools that are less relevant for the study:
\begin{itemize}
\item tools designed for citizens and non-CSIRT organisations (e.g., malware removal tools, ransomware decryptors, cyber security awareness and educational tools);

\item tools for purposes not directly related to cyber incident responses (e.g., programming tools and standard OS commands, administrative tools, internal information management, HR and finance systems);

\item tools used mainly by cybercriminals, even if they are sometimes used by CSIRTs for security evaluation and research purposes because there are too many of such tools (e.g., every single piece of malware can be considered such a tool); and

\item tools that do not have a software component (e.g., procedures, conceptual frameworks, recommendations and standards, guidelines and policies, pure hardware solutions); and

\item tools mentioned on a national CSIRT's web page without explicit evidence that a specific CSIRT used or is using it.
\end{itemize}

Additionally, shareware and commercial tools with a free option (with limited features or quota) were included but excluded commercial tools that will expire after a defined trial period. The above criteria were applied to the first and second stages to identify relevant data items.

In the following, different aspects of the procedure are explained in great detail. First, the procedure for searching into websites of national CSIRTs and cross-CSIRT organisations (Stage 1) is explained, followed by searching the research literature (Stage 2).

\subsection{Stage 1 -- Identifying Relevant Data Items on Websites of National CSIRTs and Cross-CSIRT Organisations}
\label{subsec:Stage1}

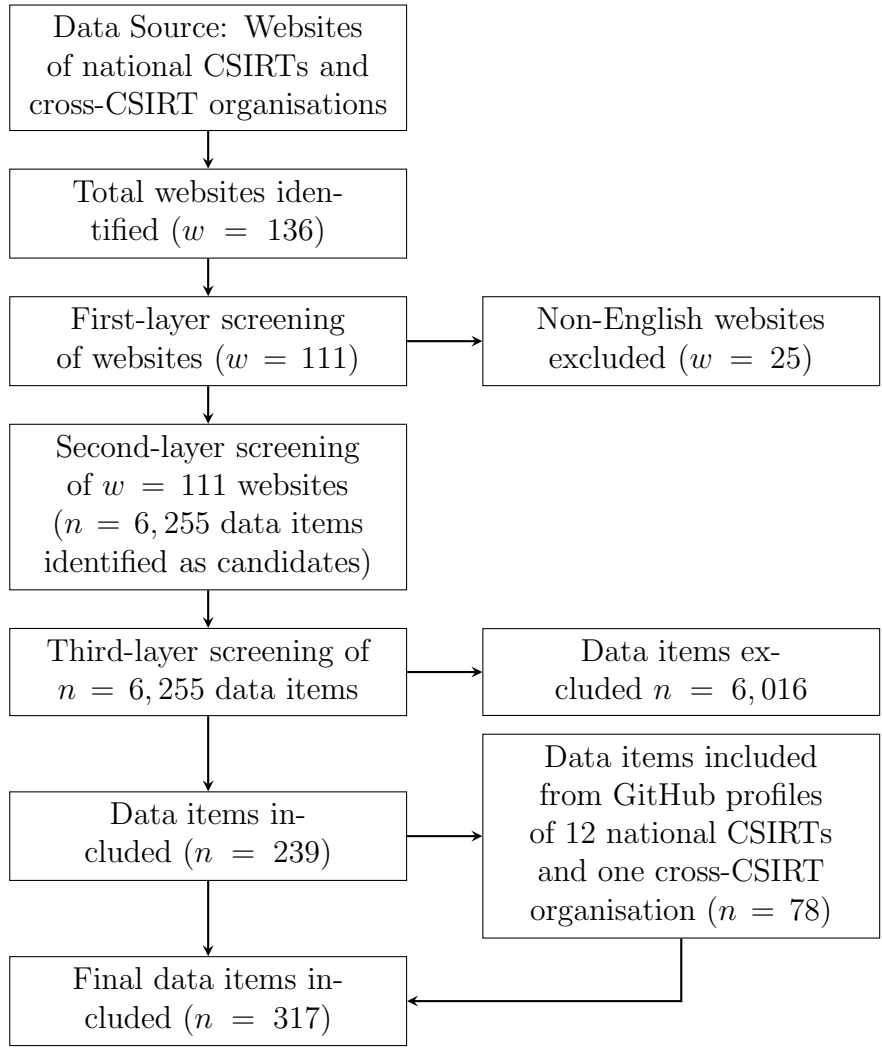
\begin{figure}[!htb]
\centering 
\begin{tikzpicture}[node distance=0.5cm]
\node (data_sources) [process] {Data Source: Websites of national CSIRTs and cross-CSIRT organisations};
\node (initial_results) [process, below=of data_sources] {Total websites identified ($w=136$)};
\node (screening_metadata) [process, below=of initial_results] {First-layer screening of websites ($w=111$)};
\node (excluded_metadata) [process, right=1cm of screening_metadata] {Non-English websites excluded ($w=25$)};
\node (included_metadata) [process, below=of screening_metadata] {Second-layer screening of $w=111$ websites ($n=6,255$ data items identified as candidates)};
\node (screening_fulltext) [process, below=of included_metadata] {Third-layer screening of $n=6,255$ data items};
\node (excluded_fulltext) [process, right=1cm of screening_fulltext] {Data items excluded $n=6,016$};
\node (included) [process, below=1cm of screening_fulltext] {Data items included ($n=239$)};
\node (adhoc) [process, right=1cm of included] {Data items included from GitHub profiles of 12 national CSIRTs and one cross-CSIRT organisation ($n=78$)};
\node (final) [process, below=1cm of included] {Final data items included ($n=317$)};
\draw [arrow] (data_sources) -- (initial_results);
\draw [arrow] (initial_results) -- (screening_metadata);
\draw [arrow] (screening_metadata) -- (excluded_metadata);
\draw [arrow] (screening_metadata) -- (included_metadata);
\draw [arrow] (included_metadata) -- (screening_fulltext);
\draw [arrow] (screening_fulltext) -- (excluded_fulltext);
\draw [arrow] (screening_fulltext) -- (included);
\draw [arrow] (included) -- (adhoc);
\draw [arrow] (included) -- (final);
\draw [arrow] (adhoc) |- (final);
\end{tikzpicture}
\caption{The procedure we followed for Stage 1 of the SLR, by searching into websites of national CSIRTs and cross-CSIRT organisations}
\label{fig:Review_Process_CSIRTPublication}
\end{figure}

The procedure for Stage 1 is shown in Figure~\ref{fig:Review_Process_CSIRTPublication}, where $w$ represents the number of national CSIRTs and cross-CSIRT organisations whose websites were searched into, and $n$ represents the number of data items -- the web pages and online documents. This stage can be further split into three steps: 1) identifying and screening the data sources (websites of national CSIRTs and cross-CSIRT organisations)--($w$), 2) searching into the websites (data sources) using the selected keywords to identify relevant data items--($n$) and 3) reading in full the data items to identify only relevant data items--($n$). The search in Stage 1 was conducted between 19 and 22 August 2021.

The SLR focused on the publicly searchable part of the official websites of national CSIRTs and cross-CSIRT organisations and identified 125 national CSIRTs from the lists provided by three key organisations: the CERT Division of the Software Engineering Institute (SEI) of Carnegie Mellon University (CMU) in the USA~\cite{Carnegie2023}, FIRST~\cite{FIRSTmembers}, and ITU~\cite{ITUnationalCIRT2022}.

The SLR also included 11 cross-CSIRT organisations' websites based on the personal knowledge of the first author's experience as the staff of a national CSIRT, a Google search with relevant keywords and checking websites of the included national CSIRTs. ENISA was included because its website has comprehensive CSIRT and digital forensic investigative tools with techniques in its training manuals for national CSIRTs.

After the first step of screening 136 websites, 25 non-English websites were excluded and included websites of 100 national CSIRTs and 11 cross-CSIRT organisations to search for relevant data items. The list of websites included for the SLR is available in the Appendix. The SLR also paid attention to all the 111 organisations and searched if they had an official profile on GitHub (\url{https://github.com/}). This led to 13 GitHub profiles owned by 12 national CSIRTs and one cross-CSIRT organisation. A Google search was not applied to such GitHub profiles since searching the profiles directly at GitHub was more accurate and easier. All public projects of those profiles were checked to identify relevant public data and free tools developed by the national CSIRT community for national CSIRTs.

\subsubsection{Keyword Selection}

The following keywords were used on Google Search to look for relevant data items (web pages and online documents) on each of the 111 websites covered. The ``[domain name]'' is replaced with the domain name of each website, removing ``www'' when relevant) as in the below search query:

\begin{quote}
\tt
site:[domain name] "open data" OR "public data" OR "free data" OR tool* OR system* OR software OR service*
\end{quote}

For some websites, the domain name is too broad that the above keyword returned too many pages, so we added the corresponding acronym of CSIRT (e.g., ``CERT'' for national CSIRTs whose names use ``CERT'') as a new keyword to limit the returned results.

For each search, all pages returned by Google were manually checked until the following standard message:
\begin{quote}
``In order to show you the most relevant results, we have omitted some entries very similar to the [X numbers] already displayed.''
\end{quote}

For each returned web page, the partial text returned by Google was read to judge if the web page or online document is likely relevant. If so, the web page or the online document is opened to read in full to check if there is any mention of the use of public data or free tools in the operational practices of one or more CSIRTs.

Links on web pages that are not hosted by the corresponding CSIRT but include some relevant information on public data and free tools were considered candidate data items. Going this additional mile is particularly important for web pages that show a list of public datasets and free tools, e.g., open data at the CIRCL (the Computer Incident Response Team Center Luxembourg, \url{https://www.circl.lu/opendata/}). Such links include GitHub profiles of national CSIRTs and cross-CSIRT organisations.

\subsubsection{Exclusion and Inclusion Criteria}

A rigid set of exclusion and inclusion criteria was not defined because some web pages and online documents only have a small section referring to the use of public data and free tools.

The following exclusion criteria (EC) were used for the study:
\begin{itemize}
\item Websites written in non-English were excluded since English is the main language the authors and readers of this SLR can understand.

\item Web pages that do not discuss or mention the use of public data or free tools by CSIRTs were excluded.

\item Web pages that discuss general guidelines, recommendations and tools for the general public and SMEs, and general security tools for pen-testing and security-testing purposes were excluded.

\item Web pages with too many secondary links, e.g., index pages, were excluded. This is because relevant pages on such pages were normally returned by Google search separately.

\item Web pages that mention common public databases such as CVE (Common Vulnerabilities and Exposures, \url{https://cve.mitre.org/}), CWE (Common Weakness Enumeration, \url{https://cwe.mitre.org/}) and NVD (National Vulnerability Database of NIST, \url{https://nvd.nist.gov/})  were excluded. This is because such data sources frequently appear on too many web pages.
\end{itemize}

After applying the exclusion criteria, the following inclusion criteria (IC) were applied to select data items for the SLR:

\begin{itemize}
\item Data items that discuss, advocate and mention the use of public data or free tools within national CSIRTs.

\item Data items about tools developed by national CSIRTs, their staff, and those tools whose descriptions show relevancy to cyber threat intelligence and cyber incident responses (e.g., recommended by a national CSIRT or a cross-CSIRT organisation for such purposes).
\end{itemize}

\subsection{Stage 2 -- Identifying Relevant Research Papers}
\label{subsec:Stage2}

\begin{figure}[!htb]
\centering 
\begin{tikzpicture}[node distance=0.5cm]
\node (data_source) [process] {Data Source: Scopus};
\node (initial_results) [process, below=of data_sources] {Articles identified ($n=577$)};
\node (screening_metadata) [process, below=of initial_results] {Screening papers based on title, abstract and keywords};
\node (excluded_metadata) [process, right=1cm of screening_metadata] {Papers excluded ($n=696$)}; 
\node (included_metadata) [process, below=of screening_metadata] {Papers included ($n=81$)};
\node (screening_fulltext) [process, below=of included_metadata] {Selecting papers based on fulltext};
\node (excluded_fulltext) [process, right=1cm of screening_fulltext] {Papers excluded ($n=59$)};
\node (included) [process, below=1cm of screening_fulltext] {Papers included ($n=22$)};
\node (adhoc) [process, right=1cm of included] {New papers identified in bibliographies of included papers ($n=2$)};
\node (final) [process, below=1cm of included] {Final selected papers ($n=24$)};
\draw [arrow] (data_source) -- (initial_results);
\draw [arrow] (initial_results) -- (screening_metadata);
\draw [arrow] (screening_metadata) -- (excluded_metadata);
\draw [arrow] (screening_metadata) -- (included_metadata);
\draw [arrow] (included_metadata) -- (screening_fulltext);
\draw [arrow] (screening_fulltext) -- (excluded_fulltext);
\draw [arrow] (screening_fulltext) -- (included);
\draw [arrow] (included) -- (adhoc);
\draw [arrow] (included) -- (final);
\draw [arrow] (adhoc) |- (final);
\end{tikzpicture}
\caption{The procedure we followed for Stage 2 of the SLR, on research papers.}
\label{fig:Review_Process_ResearchPaper}
\end{figure}
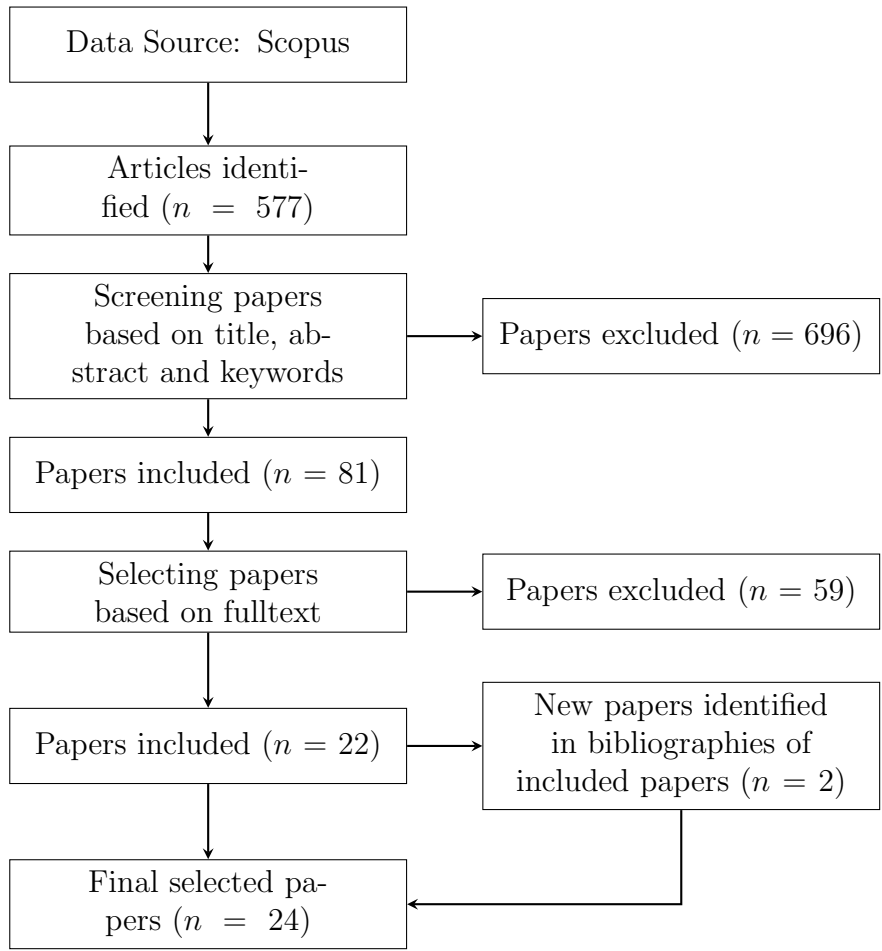

The procedure for this stage is depicted in Figure~\ref{fig:Review_Process_ResearchPaper}. A systematic literature search was conducted between 17 and 18 November 2022 on Scopus -- one of the most widely used scientific databases of research papers -- by applying a keyword search into metadata (titles, abstracts and keywords).

After applying the exclusion and inclusion criteria, 22 relevant research papers were identified ~\cite{Kassim2022, riebeimpact, haidar2021analysis, hossain2021cyber, Hellwig2018GDPR, Jaatun2020, Kashiwazaki2018, Kijewski2011, Konno2018, Krstic2019, Leppanen2020, Ma2012, mana2019eurocontrol, Metzger2011, Mokaddem2019, Mooi2016, Oguchi2018, Reyes2018, ruefle2014computer, Skopik2016, Valladares2017, Yamin2019}. Then we looked into the bibliographies of the 22 papers and identified one more research paper~\cite{Kampanakis2014} and a technical report~\cite{Bourgue2013} that was not covered by the keyword search in Scopus. Finally, we identified 23 research papers and one technical report for the SLR, leading to 24 data items for Stage 2.

\subsubsection{Data Source}

For the data source, Scopus, one of the most used and popular scientific indexing databases for searching the research literature, was selected for the study~\cite{Zhu2020}. The Web of Science (WoS) database was also considered, but a search with the Scopus database returned more results than the Web of Science (WoS) database. This is unsurprising when referring to previous comparative studies of different scientific databases~\cite{aghaei2013comparison, Pranckute2021WoS-vs-Scopus, Visser2021Scopus-vs-others}. Furthermore, searching into specific scientific publishers' databases was refrained because Scopus already has very good coverage of the most important scientific publishers to answer the RQs of the study. A search with Google Scholar was attempted but was not considered because it lacked the support of advanced search syntax, such as nested searches and digging into metadata. As an automated search engine, it was found Google Scholar provides less accurate search results and covers too many data items, with papers that are not peer-reviewed and published at problematic venues (e.g., predatory journals and conferences). Google Scholar also returns a limited number of relevant papers, making it less ideal even if more complete coverage is needed\footnote{See \url{https://harzing.com/resources/publish-or-perish/manual/reference/dialogs/preferences-google-scholar} for some explanation on this search limitation.}

\subsubsection{Search Keywords}

To determine the best search query, various combinations of different keywords were tried and judging the returned articles, the search query was finally decided. This ensured covering as many relevant papers as possible while returning a manageable number of candidates for subsequent manual screening. The search terms ``security OR attack*'' were used with ``CSIRT'' and ``CERT'' acronyms to eliminate papers that use identical acronyms but are unrelated to our study. The final search query used is shown below:
\begin{center}
\tt
((cyber* OR computer) AND (incident* OR emergenc* OR security OR attack*) AND ("response team*" OR "readiness team*")) OR ((CSIRT OR CSIRTs OR CSIRT OR CSIRTs OR CERT OR CERTs) AND (security OR attack*))
\end{center}

\subsubsection{Exclusion and Inclusion Criteria}

The following EC were used to screen research papers:

\begin{itemize}
\item Papers published before 2010 were excluded to ensure that the papers included reflect the more recent and currently valid practices of national CSIRTs.

\item Non-English papers were excluded since English is the primary language the authors and readers of this SLR can understand.

\item Papers that are not discussing  CSIRTs operations were excluded.

\item Papers that do not report original work/opinions (e.g., editorials) were excluded.

\item Papers with unclear abstracts (in terms of the quality of technical writing) were excluded because unclear or poorly written abstracts normally imply that the results reported are unreliable or difficult to interpret.
\end{itemize}

After applying the EC, the following IC was applied to select the most relevant papers for the SLR:

\begin{itemize}
\item papers that cover operational practices in CSIRTs concerning public data or free tools,

\item papers that report empirical studies, literature surveys and systematic literature reviews regarding the use of public data or free tools in CSIRTs,

\item papers that report cyber incident case studies using public data or free tools in CSIRTs, and

\item papers that propose new approaches to enhance operational practices and capabilities of CSIRTs with the use of public data or free tools.
\end{itemize}

\subsection{Data Analysis}

A mixed approach, qualitative and quantitative, was used to review the contents of the documents extracted from national CSIRT websites, cross-CSIRT websites and research papers identified from a scientific database. A quantitative approach was used to get an overview of the documents and papers in the form of numbers, e.g. counting the number of papers published by year and illustrated in figures. Qualitative analysis was used to generate the main theme, keeping in mind the Research Questions from which the main findings for the study are drawn to answer the Research Questions. 


The content analysis method was used in previous studies conducting systematic literature reviews~\cite{germain2010influence, singh2019many, ahmad2020systematic} that motivates us to adopt a similar method. \emph{Content analysis} is considered the best method to generate findings from the research papers and put them into the context of the research questions~\cite{white2006content}. Additionally, content analysis is a flexible approach~\cite{cavanagh1997content, tesch2013qualitative}, yet systematic, rigorous~\cite{white2006content} and suitable when an existing theory or research literature on a particular topic is limited~\cite{hsieh2005three}, exemplified by the present study.

\section{Results} 
\label{sec:Results-SLR}

This section presents Stage 3, the synthesis of results from Stage 1 (data items from National CSIRT websites and cross-CSIRT organisations) and Stage 2 (research papers). Overall, in Stage 1,  317 data items were identified from the websites of 100 national CSIRTs and 11 cross-CSIRT organisations and in Stage 2, 24 research publications relevant to the study were identified from the Scopus scientific database.

The result from the SLR are presented under the following themes, with the RQs in mind; 1) General information, 2) Use of public data in national CSIRTs, 3) Use of free tools in national CSIRTs, 4) Active development and advocating of public data and free tools by national CSIRTs and Cross-CSIRT organisations, 5) How national CSIRTs' staff perceives public data and free tools, 6) Information sharing among national CSIRTs and 7) Challenges and issues with tools and data in national CSIRTs' operational practices.   



\subsection{General Information}

\subsubsection{Year of Publication}

The publication year is not always known for the identified web pages and online documents, so their statistics are not shown. Nevertheless, our manual inspection revealed that most web pages and online documents were published in the past five years.

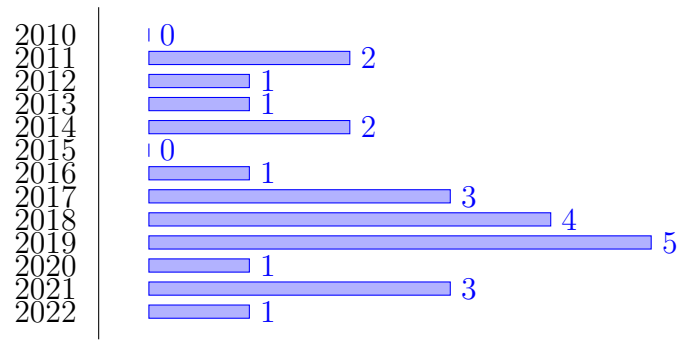
\begin{figure}[!htb]
\centering
\begin{tikzpicture}
\begin{axis}[
width=0.7\linewidth,
height=6cm,
xbar,
bar width=5pt,
axis lines*=left,
axis x line=none,
tick style={draw=none},
ytick=data,
y dir=reverse,
y tick label style={/pgf/number format/.cd,scaled y ticks = false,set thousands separator={},fixed},
nodes near coords, 
nodes near coords align={horizontal},
]
\addplot coordinates {(0,2010) (2,2011) (1,2012) (1,2013) (2,2014) (0,2015) (1,2016) (3,2017) (4,2018) (5,2019) (1,2020) (3,2021) (1,2022)};
\end{axis}
\end{tikzpicture}
\caption{Number of research publications identified by year}
\label{figure:papers_by_year}
\end{figure}

Similarly, most research publications were published within the last five years, as shown in Figure~\ref{figure:papers_by_year}.

\subsubsection{Types of CSIRTs}

While the CSIRTs' publications covered national CSIRTs only, the 24 research publications also covered the different types of CSIRTs, as shown in Table~\ref{table:Types_CSIRTs}. This is due to the lack of research publications on national CSIRTs. Hence, we include a list of publications that mention other CSIRTs to get a general understanding of CSIRTs. Simultaneously, this provides knowledge about the different types of CSIRTs in the research literature.

\begin{table}[!htb]
\caption{Different types of CSIRTs mentioned in research publications}
\label{table:Types_CSIRTs}
\centering
\begin{tabular}{ll}
\toprule
\textbf{Type of CSIRTs} & \textbf{Research Papers} \\
\midrule
Regional CSIRTs & \cite{Kassim2022}\\
National CSIRTs & \cite{Leppanen2020, Mokaddem2019, Bourgue2013, Kassim2022,hossain2021cyber, haidar2021analysis, riebeimpact}\\
Public Sector CSIRTs & \cite{riebeimpact}\\
Organisational CSIRTs & \cite{Konno2018, mana2019eurocontrol, Oguchi2018, Yamin2019, Skopik2016, Jaatun2020, Krstic2019, Ma2012}\\
Research Organisation CSIRTs& \cite{Metzger2011, Kijewski2011}\\
Academic CSIRTs& \cite{Kashiwazaki2018, Reyes2018}\\
\bottomrule
\end{tabular}
\end{table}

\subsubsection{Summary of Research Papers Covered in the SLR}

The research papers cover various topics about CSIRTs, adopting different research methods. The details of the 24 research publications covered in the SLR are summarised in Table~\ref{table:Method_research}.

\begin{longtable}{p{0.09\linewidth} p{0.65\linewidth} p{0.15\linewidth}}
\caption{Summaries of the 20 research publications covered in the SLR}\label{table:Method_research}\\
\toprule
\textbf{Cite} & \textbf{Research Topic} & \textbf{Research Method}\\
\midrule
\endfirsthead

\multicolumn{3}{c}%
{\tablename\ \thetable\ -- \textit{Continued from previous page}} \\
\toprule
\textbf{Cite} & \textbf{Research Topic} & \textbf{Research Method} \\
\midrule
\endhead

\endfoot

\bottomrule
\endlastfoot

\cite{Kassim2022}& The lack of information about national CSIRTs, the systematic use of data and tools and the evaluation of tools and data need improvement. & Case Study\\
\midrule
\cite{riebeimpact}& The need for improvement in the current operational practices, e.g. analysis of incident data, transparent reporting and tool structures for information exchange. & Empirical Study\\
\midrule
\cite{hossain2021cyber}& A proposal to address the lack of detection tools, incident response tools and procedures governing the operations of national and other types of CSIRTs. & Case Study\\
\midrule
\cite{haidar2021analysis}& Examine national CSIRTs in Indonesia using the Carnegie Mellon University framework and proposed recommendations for improvement based on the FIRST framework. & Case Study\\
\midrule
\cite{Leppanen2020} & To examine citizens' and organisations' roles within the formal cyber security network in identifying and reporting incidents in a country. & Nodal cluster analysis\\
\midrule
\cite{Jaatun2020} & To investigate cyber incident response readiness and capacity in the Norwegian petroleum industry for handling critical ICT security incidents. & Empirical study'\\
\midrule
\cite{Krstic2019} & The study investigates the applications of machine learning in cyber security and how machine learning can be adopted to improve operations in CSIRTs. & Literature survey\\
\midrule
\cite{mana2019eurocontrol} & Describing the European Air Traffic Management CERT's functions and services and partnerships with other CSIRTs on threat information sharing to detect, respond, mitigate and prevent cyber threats. & Case study\\
\midrule
\cite{Mokaddem2019} & Development of an automated tool, the AIL Framework to detect information leaks in Luxembourg, with contextual information by crawling paste websites for incident response escalation. & Tool development\\
\midrule
\cite{Yamin2019} & To investigate the practice of automatic information exchange among CSIRTs, the tools, approaches, standards used, challenges and technologies for automating information exchange. & Systematic literature review\\
\midrule
\cite{Kashiwazaki2018} & The study is about responding to an information leak incident at a University by the university's CSIRT. & Case study\\
\midrule
\cite{Konno2018} & Integrating cyber threat intelligence in corporate-based security frameworks, such as NTT CSIRT, to defend networks and systems against sophisticated cyber attacks. & Case study\\
\midrule
\cite{Oguchi2018} & A successful forensic-based incident response in NEC CSIRT by using an in-house developed tool to preserve  data and evidence. & Case study\\
\midrule
\cite{Hellwig2018GDPR} & A GDPR compliance checking model and tool for supporting the information exchange between CSIRTs, to be integrated into the current CSIRTs' information sharing process. & Model and tool development\\
\midrule
\cite{Reyes2018} & Automating the dissemination of information from an Academic CSIRT for real-time identification of possible new cyber security threats and cyber-attacks. & Business Intelligence applications, Extract, Transform and Load and Online Analytical Processing with Ralph Kimball methodology.\\ 

\midrule
\cite{Valladares2017} & Development of early warning alerts of upcoming malicious activities for CSIRTs to improve overall security. &  A Data Warehouse and Business Intelligence applications, with Ralph Kimball methodology.\\

\midrule
\cite{Mooi2016} & Designed a model for establishing CSIRTs, to address  the deficiency of a standard model for developing CSIRT & Design Science Research\\ 

\midrule
\cite{Skopik2016} & A critical review of state of the art and essential considerations when building effective security information sharing platforms for CSIRTs. & Literature survey\\
\midrule
\cite{ruefle2014computer} & Establishment of a CSIRT, its purpose of dealing with attacks, evaluating CSIRTs' performance and the information sharing as a core service. & Case study\\
\midrule
\cite{Kampanakis2014} & Describes and discusses a list of options on information sharing for CSIRTs, its common uses, and clarifies misconceptions about data sharing. & General discussions\\
\midrule
\cite{Bourgue2013} & Investigate existing communication solutions and practices among European CSIRTs, and gaps that limit threat intelligence exchange in Europe. & Literature survey and interviews\\
\midrule
\cite{Ma2012} & Presents Nagios for monitoring known vulnerabilities and Pakiti for systematic patches to vulnerable systems in EGI CSIRT. & Case study\\
\midrule
\cite{Kijewski2011} & Introduces in-house projects and systems for automatically detecting threats such as malware infections in Poland. & Case study\\
\midrule
\cite{Metzger2011} & Orchestration of reporting capabilities, automatic analysis and response, and process-oriented intervention for integrated management of security incidents & Case study\\
\end{longtable}

\subsection{Use of Public Data in National CSIRTs' Operations}
\label{subsec:Public_Data}

More than three-quarters (263) of the web pages and online documents identified in Stage 1 mentioned free tools compared to public data. To further understand if these free tools utilise public data, we looked into the respective websites of all free tools mentioned in the identified web pages. We found that some tools use a comprehensive list of public data, e.g., IntelMQ, which reads from many public data sources (\url{https://intelmq.readthedocs.io/en/latest/user/feeds.html}) while others do not.

We identified various public data used by national CSIRTs, who create and maintain some of these public data. These data can be classified as below:

\begin{itemize}
\item \textbf{Data about known vulnerabilities and exposure}: CIRCL CVE daily JSON dump (\url{https://cve.circl.lu/static/circl-cve-search-expanded.json.gz})

\item \textbf{Security feeds and threat intelligence data}: CIRCL Images Phishing Dataset (\url{https://www.circl.lu/opendata/circl-phishing-dataset-01/}), CIRCL Images AIL Dataset (\url{https://www.circl.lu/opendata/circl-ail-dataset-01/}), IOC-DB (\url{https://labs.inquest.net/iocdb}) and MITRE ATT\&CK (\url{https://attack.mitre.org/versions/v8/software/});

\item \textbf{Data about domain registration information}: WHOIS Data (\url{http://https://lookup.icann.org/}), RIPE  database (\url{https://www.ripe.net/manage-ips-and-asns/db}), CERT.at Taxonomy of Domain Names Labelling (\url{https://github.com/certat/awesome-taxonomyzoo-list}).

\item \textbf{General-purpose data}: Government of Serbia open data (\url{https://www.ite.gov.rs/tekst/en/30/open-data-portal.php}), Latvia institutions open data (\url{https://data.gov.lv/eng/about});

\end{itemize}

National CSIRTs rely on data from various sources and a wide range of tools for daily incident responses~\cite{Kassim2022}. National CSIRTs use closed-source and public data to help enrich information about threats and facilitate incident responses. Commonly used \emph{closed-source data} sources include incident reports, cyber threat intelligence data and security feeds provided by various organisations (e.g., Shadowserver for taking down botnets and malicious sites). \emph{Public data} mainly refers to data publicly available on the Internet, often obtained via Open Source Intelligence (OSINT) tools.

Not many research papers from the literature have comprehensive coverage concerning public data in national CSIRTs and CSIRTs, in general. Several (8) research papers did indicate that public data is used in CSIRTs' operations, but information about what type of public data and how it is used and perceived is largely missing. The public data identified from research publications can be conceptually categorised as follows:

\begin{itemize}
\item \textbf{Data about known vulnerabilities and exposure}: Common Vulnerabilities and Exposures (CVE, \url{https://cve.mitre.org/})~\cite{Mokaddem2019, Ma2012}, National Vulnerability Database (NVD, \url{https://nvd.nist.gov/})~\cite{Kashiwazaki2018}, 

\item \textbf{Security Feeds and Cyber Threat Intelligence datasets}: Team Cymru Hash Registry (\url{https://team-cymru.com/community-services/mhr/})~\cite{Bourgue2013}, data from Malware Capture Facility Project (\url{https://mcfp.weebly.com/})~\cite{Reyes2018}, Spamhaus blocklists (\url{https://www.spamhaus.org/})~\cite{Bourgue2013}, Zone-H.org (a crowdsourcing-based dataset of defacement attacks)~\cite{Valladares2017}, malwaredomains.com~\cite{Kijewski2011}, Shadowserver(\url{https://www.shadowserver.org/})~\cite{riebeimpact, Kassim2022};

\item \textbf{Data on general-purpose public websites}: public media news~\cite{Leppanen2020}, posts on public web forums~\cite{Skopik2016}, data on online social networks (OSNs) such as Twitter~\cite{Kashiwazaki2018}, TweetDeck~\cite{riebeimpact},  Manufacturer websites~\cite{riebeimpact};

\item \textbf{Data on public websites with a strong ICT flavour, not for cyber security per se only}: data shared on websites such as Pastebin.com~\cite{Mokaddem2019, Konno2018}, gist (\url{https://gist.github.com/})~\cite{Mokaddem2019} and codepad (\url{http://codepad.org/})~\cite{Mokaddem2019}, data on Alexa.com~\cite{Kijewski2011}, Security Advisories~\cite{riebeimpact};
 
\item \textbf{Publicly accessible data on darknet}~\cite{Konno2018, Mokaddem2019}.

\item \textbf {Open-source intelligence (OSINT)}~\cite{riebeimpact, Kassim2022};

\item \textbf {Cyber range data}~\cite{hossain2021cyber};
\end{itemize}

The above lists are neither complete nor representative of national CSIRTs due to the small number of research papers covered by our SLR and the ad hoc mentions of public data in the national CSIRT and cross-CSIRT websites. It appears that national CSIRTs may not have utilised some public data sources due to a lack of systematic information about such data and the need to search them easily~\cite{Kassim2022}. Therefore, more studies into this topic are suggested.



\subsection{Use of Free Tools in National CSIRTs Operations}
\label{subsec:Free_Tools}

Publications from national CSIRTs, Cross-CSIRT organisations' and their respective GitHub websites mentioned 301 free tools, and the research papers mentioned 64 free tools. The CSIRT publications and research papers mentioned include free tools, open-source tools, closed-source tools, freeware and free online services. The free tools mentioned in CSIRT publications can be classified based on the following tasks: Log Analysis, Artefact Analysis, Incident Management, Whois, Web Information Gathering, Network Monitoring, Sandbox and Reverse Engineering, Penetration Testing, Parser, Scanner, Disk Image Creation, Evidence Collector, Visualisation, Cyber Threat Intelligence tool, Honeypot, Memory Analysis tool, Search Engine, File Recovery, Linux Distributions, Communication tools, Network Utility and Intrusion Detection System.

A comprehensive list of free tools, identified through a systematic search into national CSIRTs' websites, Cross-CSIRT organisations and research publications, is available at:

\url{https://cyber.kent.ac.uk/research/CSIRTs/List_Open-source_Free_Tools.html}

The study found the developers of these free tools consist of software companies, researchers affiliated with research organisations and CSIRTs, independent researchers, independent developers, CSIRTs staff and joint projects or collaborations between CSIRTs and researchers. Besides being used in national CSIRTs' operations, the tools are also recommended for all types of CSIRTs and the wider security practitioners. 

A number of national CSIRTs actively develop tools for operational use and share them with others. IntelMQ, developed by CERT.at (Austria) and used by BGD eGOV CIRT (Bangladesh) exemplify this. ENISA mentioned that several national CSIRTs use IntelMQ across Europe, such as SWITCH-CERT (Switzerland) and NCSC-NL (Netherlands). Other free tools developed by national CSIRTs include Assemblyline (cyber security Centre of Canada), Gabriel (Georgia's national CSIRT), IFAS.io (Hong Kong national CSIRT), DNSCheck (Iran national CSIRT), TARANIS (NCSC-NL) and  Pastelyzer (national CSIRT of Latvia). Other tools developed by national CSIRTs can be referred at Table \ref{table:nationalCSIRT_freetools}. National CSIRTs also use third-party free tools to support operations, such as the US-CERT (USA) uses YARA to detect malware, CIRCL uses TheHive for investigating cyber incidents and several CSIRTs across Europe use AIRT for incident tracking.

Meanwhile, tools mentioned in the research papers cover those that support incident response (investigative tools, ticketing systems, reporting systems), intrusion detection systems, information sharing (information exchange) platforms and OSINT tools in CSIRTs. Other tools are those developed by the authors of the research papers that CSIRTs and security practitioners can also use. The authors use a small number of tools mentioned to facilitate their own research.  
Tools identified from the research papers can be classified into different classes, shown with the name of tools in Table \ref{table:Class_tools}. 

\begin{table}[!htb]
\caption{Sample Tools Identified from Research Papers}
\label{table:Class_tools}
\centering
\begin{tabular}{ p{0.38\linewidth} p{0.53\linewidth} }
\toprule
\textbf{Class of Tools} & \textbf{Example Tools with Citation} \\
\midrule
SIEM & iView~\cite{Skopik2016, hossain2021cyber}, OSSIM~\cite{Metzger2011}, SIEM ~\cite{hossain2021cyber}\\
Vulnerability Scanners & nmap~\cite{Metzger2011}, deep exploit~\cite{Krstic2019}, OpenVAS~\cite{Kashiwazaki2018}\\
Information Sharing & MISP~\cite{Mokaddem2019, Yamin2019, Skopik2016, Hellwig2018GDPR, mana2019eurocontrol, riebeimpact}, CIF~\cite{Skopik2016}, TAXII~\cite{Skopik2016,Hellwig2018GDPR}\\
Incident Management & RTIR~\cite{Mokaddem2019}\\ 
Incident reporting & Email, telephone~\cite{riebeimpact}\\
TheHive~\cite{Mokaddem2019, Bourgue2013}, OTRS~\cite{riebeimpact}\\
Network Monitoring & Autoreport~\cite{Leppanen2020}, Nyx~\cite{Metzger2011}, Nagios~\cite{Ma2012}\\
Artefact analysis  & Cuckoo sandbox~\cite{Krstic2019}, Nfdump~\cite{Metzger2011}, LogHound~\cite{Skopik2016}, YARA~\cite{Kampanakis2014}\\
Communication & Email~\cite{Bourgue2013, riebeimpact}, Skype~\cite{Bourgue2013}, IRC~\cite{Jaatun2020}\\
Intrusion Detection System & Snort~\cite{Reyes2018, Metzger2011}, IDS~\cite{riebeimpact}\\
Web Information Gathering & AIL framework ~\cite{Mokaddem2019}\\
Social Media Monitoring & ScatterBlogs~\cite{riebeimpact}\\
Communication channels & Chat~\cite{riebeimpact}, Wiki page~\cite{riebeimpact}\\
Threat Intelligence & OSINT Tool~\cite{Kassim2022}\\

\bottomrule
\end{tabular}
\end{table}

Several research publications had pointed out the significance of tools and data to facilitate incident responses in organisational CSIRTs' operations, not precisely in national CSIRTs. For instance, Toendel et al.~\cite{Tondel2014}conducted a systematic literature review of incident response practices of various organisations. The authors highlighted the importance of tools support and automation that are vital to facilitate incident response. The systematic literature review findings by Toendel et al. are similar to a case study by Kijewski et al.~\cite{Kijewski2011} who examined the operations of CERT Polska (a CSIRT under the Research and Academic Institution). The authors revealed the importance of tools and data used in CERT Polska, e.g., threat detection tools, honeynets, and classified and closed-source data, to respond to and detect cyber threats more efficiently and timely. 

Mana and Vasileios's~\cite{mana2019eurocontrol} experiences and personal observation provide a better understanding of the active use of an automated tool such as Malware Information Sharing Platform (MISP) (\url{https://www.misp-project.org/}) in the operational practices at Eurocontrol Air Traffic Management CSIRT (EUROCONTROL/EATM-CERT). The authors also revealed that threat data and other threat information needed for incident response are actively shared among internal CSIRTs within the EATM-CERT and with several national CSIRTs in Europe, benefiting from these tools.

Metzger et al.~\cite{Metzger2011} reported through a case study of the practices at the Leibniz Supercomputing Centre (LRZ-CSIRT), an Academic CSIRT in Germany. The authors provided a good understanding of the current approach in incident response at LRZ-CSIRT consists of the orchestration of reporting capabilities, automatic analysis and response, and process-oriented intervention for integrated management of security incidents. 

Hossain et al.~\cite{hossain2021cyber} highlighted on Security Information \& Events Manager (SIEM) application as essential for  monitoring and incident responses for national CSIRTs. SIEMs can be used to display the data collected by network and security systems to visualise and prioritise the workload in national CSIRTs. The authors also made clear the importance of a cyber range for simulating data and security attacks against, for instance, critical infrastructure targets, as a learning and process improvement in national CSIRTs. 

Riebe et al.~\cite{riebeimpact} highlighted the use of OTRS as the ticketing system, email and telephone as means for incident reporting in public sector CSIRTs in Germany. Notably, excel sheets were used in the past to manage incidents in public sector CSIRTs in Germany. IDS are also used in German's public sector CSIRTs to detect and identify incidents, while MISP is used for threat intelligence sharing and the tool ScatterBlogs is used for monitoring and analysing social media information.

A noteworthy point highlighted by all the authors above is the need for constant progress in security-related research, which calls for more research into this area with attention specifically related to tools and data. As our lives are getting increasingly dependent on the Internet and information systems, network threats have a huge potential for impact as well as being more widespread, necessitating constant progress in security-related research~\cite{Kijewski2011}. 

\subsection{Active Development and Advocating of Public Data and Free Tools by National CSIRTs and Cross-CSIRT Organisations}
\label{CSIRT_activity}

Interestingly, the study found a few national CSIRTs who are active in developing and promoting public data and free tools, reflected from the data items identified in the SLR, summarised in Table \ref{table:nationalCSIRT_freetools} and Table \ref{table:nationalCSIRT_publicdata}. CIRCL is one of them, indicating a team from a small country that significantly impacts the CSIRT community. 

\begin{table}[!htb]
\caption{Sample Tools Developed by National CSIRTs}
\label{table:nationalCSIRT_freetools}
\centering
\begin{tabular}{ p{0.38\linewidth} p{0.53\linewidth} }
\toprule
\textbf{National CSIRTs} & \textbf{Tools Developed} \\
\midrule
CIRCL (Luxembourg) & MISP, AIL Framework, Circlean, BGPranking, URL-abuse,Potiron, Carl-Hauser, D4 Attack Map, DEFT, Douglas-Quaid \\

CERT.at (Austria) & IntelMQ, CERTspotter-processing, IP2nat, FollowTcpStream \\ 
JPCERT/CC (Japan) & MalConfScan , aa tool, Sysmon Search, Logon Tracer, Emocheck, MalConfScan-with-Cuckoo, DetectLM  \\

INCIBE-CERT (Spain) & BotChecker Script, Terminology extractor, Merovingio, Onion Indexer\\  
US-CERT (USA) & CHIRP IOC Detection Tool, Sparrow, Aviary \\ 

\bottomrule
\end{tabular}
\end{table}

Additionally, many national CSIRTs also actively recommend various third-party tools to the CSIRT community to facilitate incident responses. Additionally, national CSIRTs also develop repositories of public datasets and advocate for the CSIRT community and the public, as shown in Table \ref{table:nationalCSIRT_publicdata}.

\begin{table}[!htb]
\caption{Public Dataset Developed by National CSIRT}
\label{table:nationalCSIRT_publicdata}
\centering
\begin{tabular}{ p{0.38\linewidth} p{0.53\linewidth} }
\toprule
\textbf{National CSIRTs} & \textbf{Public Dataset Developed} \\
\midrule
CIRCL (Luxembourg) & CIRCL Images Phishing Dataset,CIRCL Images AIL Dataset on Infor
mation leak,Allaple malware infection raw data\\
CERT.at (Austria) & Taxonomy of Domain Names Labelling, Internet-inventory of  metadata on IPs and networks and ASNs on the net \\ 
JPCERT/CC (Japan) & Cyber green database of cyber security risks and vulnerabilities  \\
INCIBE-CERT (Spain) & ICARO database of IOCs\\
SWITCH-CERT (Switzerland) & Connectcome Knowledge of scientific data across multi-disciplines\\

\bottomrule
\end{tabular}
\end{table}

Furthermore, security organisations also share restricted data with national CSIRTs for national-level takedowns of botnets and phishing websites. Such data is not available to the public. This is exemplified by Shadowserver Foundation (\url{https://www.shadowserver.org}) which shares data on botnet infections to national CSIRTs and other trusted entities for national-level eradication of botnets. Team-Cymru (\url{https://www.team-cymru.com}) also works with national and regional CSIRTs worldwide by sharing similar data, such as threat intelligence,  so national CSIRTs are informed of the threats in their constituencies.

Several research papers also advocate tools such as a Security Information \& Events Manager (SIEM) application, considered essential to national CSIRTs to facilitate monitoring and incident response~\cite{hossain2021cyber}. Alerts and events generated from SIEMs are important to visualise and prioritise the incidents in national CSIRTs. It is critical for national CSIRTs to identify the infrastructure and tools needed in the operations, hence prioritising the use of open technology which is free, independent and cost-saving.~\cite{haidar2021analysis}

\subsection{How Public Data and Free Tools are Perceived by National CSIRTs' Staff}

None of the data items we identified from CSIRT publications explicitly indicates how CSIRT staff perceive the usefulness of public data and free tools. However, the fact that a number of CSIRTs actively develop and advocate public data and free tools highlighted in \ref{CSIRT_activity} implies a positive attitude and perception of public data and free tools for the operations of national CSIRTs. The fact that a number of national CSIRTs (16) identified from the study who use free tools (inclusive open-source tools) indicate that free tools are well accepted and utilised in national CSIRTs' operations. Our finding is supported by another study which found the most advanced and popular tools used in the CSIRT community are open-source tools developed by national CSIRTs, and support for these tools is very strong~\cite{CSIRT2019maturity}.  

Similarly, none of the research papers has empirical evidence of comprehensive, explicit and direct coverage of how national CSIRT staff perceive public data and free tools. Nevertheless, the usage of free tools and public data~\cite{Mokaddem2019, Metzger2011, mana2019eurocontrol, Kashiwazaki2018, Oguchi2018, Kassim2022, riebeimpact} in national CSIRTs, indirectly implies the staff's sound acceptance of free tools and public data in their operations. General CSIRT staff satisfaction reflected from surveys and interviews from previous studies~\cite{Jaatun2020, Valladares2017, Ma2012, Bourgue2013} indirectly shows positiveness towards public data and free tools. 

\subsection{Information Sharing Among National CSIRTs}

It was found in national CSIRTs' websites and cross-CSIRT organisations that national CSIRTs actively collaborate and share information, threat data and free tools within the national CSIRTs' community. This includes initiatives and projects, mainly by national CSIRTs and Cross-CSIRT organisations advocating information sharing. For instance, the MeliCERTes Cyber Security Platform (MelliCERTes CSP, \url{https://github.com/melicertes/csp}), a modular platform developed by several national CSIRTs in Europe, allows national CSIRTs to share information and collaborate within a verified trust circle. The EU-supported CyberExchange project~\cite{cyberexchange2017} is another initiative for the exchange of information, knowledge and skills among national CSIRT staff in Europe. National CSIRTs around the world also collectively exchange information on indicators of compromise (IOCs) through MISP \url{(https://www.misp-project.org/)}, a platform developed together with several CSIRTs from the European region.


Cyber-attacks, that cross boundaries between countries, need international collaboration to respond and mitigate effectively, necessitates comprehensive and effective global and regional collaboration ~\cite{hossain2021cyber}. This is further supported by the importance of global cooperation among international security players on sharing cyber security information~\cite{Hellwig2018GDPR} and operational information about threats for quick detection and response of incidents~\cite{Kijewski2011}. This includes information sharing among CSIRTs, including national CSIRTs through private-public collaboration and international cooperation in a more structured manner, essential for the operations of national CSIRTs~\cite{Skopik2016, Konno2018}. This can be through closed `mailing lists' or `encrypted email', in line with pre-agreed ``standard formats'' for information sharing~\cite{Hellwig2018GDPR} and automated information exchange such as by MISP~\cite{Mokaddem2019, Yamin2019, Skopik2016, mana2019eurocontrol}, CyBOX, CYBEX, STIX and TAXII~\cite{Yamin2019, Bourgue2013, Mooi2016} between CSIRTs, national CSIRTs and security community. 

There are different protocols and standards that regulate the work, information management, and communication, including information sharing following the Traffic Light Protocol (TLP)~\cite{FIRST-TLP, US-CERT-TLP}, to determine the confidentiality of information during communication by classifying documents or information as red, amber, green, or white. The TLP is necessary when sharing information about intelligence, indicators of compromise, incident reports, vulnerability information and malware detection~\cite{mana2019eurocontrol, Yamin2019, Bourgue2013, Jaatun2020} among CSIRTs. Nevertheless, the effectiveness of current information sharing among CSIRTs, in general, is questionable~\cite{Yamin2019}, whereby technical issues were identified as the most common barrier to effective information exchange among CSIRTs in general~\cite{Bourgue2013}. 

Many national CSIRTs actively work with each other, and a number of cross-CSIRT organisations have been established to facilitate such collaboration, including FIRST, ITU, CERT Division of SEI of CMU, APCERT, ENISA, OIC-CERT and AfricaCERT~\cite{Kassim2022}. Among others, these initiatives provide training opportunities, resource sharing and invitations to participate in cross-CSIRT cyber exercises. For instance, MyCERT conducted incident response training for newly established national CSIRTs from the South East Asia region via initiatives of the APCERT. Among all cross-CSIRT organisations, FIRST is of particular importance because it has national and organisational CSIRTs as members, and it organises a wide range of initiatives at the global level~\cite{Kassim2022}. Nevertheless, it remains largely unknown how helpful such cross-CSIRT initiatives are for improving national CSIRTs' operation, hence the need for more prospective research on this topic~\cite{Kassim2022}.

\subsection{Challenges and Issues with Tools and Data in National CSIRTs' Operational Practices}
\label{subsubsec:Challenges}


Documents from national CSIRTs' websites and cross-CSIRT organisations did not highlight challenges and issues in national CSIRTs' operational practices. Nonetheless, research papers did highlight some issues and problems in the current operational practices of national CSIRTs and CSIRTs in general. Tools and data are crucial for CSIRTs to support their operations, but there are some issues and challenges with regard to the quality and usability of tools in CSIRTs. Werlinger et al.~\cite{Werlinger2010}, through an exploratory study of incident response practices in the academic, governmental, and private sectors, identified some issues with tools in the operations of organisational CSIRTs. The authors pointed out that practitioners often need to develop in-house tools for diagnostics and detection of incidents, as an incident response is highly collaborative work. Furthermore, the detection of incidents is becoming complicated, requiring expertise on the part of the staff and effective support from usable and reliable security tools. The point highlighted by Werlinger et al.~\cite{Werlinger2010} is crucial, which indicates a gap and requires attention from researchers and stakeholders. 

Similar to Werlinger et al.~\cite{Werlinger2010}, Metzger et al.~\cite{Metzger2011} identified issues concerning tools, particularly with regard to the handling of tools, precisely forensic tools and evidence collection. While Werlinger et al. revealed the issue with general incident response tools, Metzger et al. were more specific in highlighting the issue of forensic tools in the CSIRT. Nevertheless, a crucial point the authors are trying to convey is the issue and problems with tools used for investigating incidents in CSIRTs that need prospective studies to address the issues. 

Several research papers~\cite{Krstic2019, Kashiwazaki2018, Kijewski2011, Metzger2011} highlighted the operational challenges of CSIRTs in general (not precisely national CSIRTs) concerning the need for automation and advanced tools to facilitate effective detection and analysis of cyber incidents, which worth highlighting in this study. Manual handling of incidents is found to be a challenge in CSIRTs as it is prone to human errors causing loss of incident evidence~\cite{Oguchi2018}. The challenge with ``Information overload” concerning vulnerabilities and attacks in CSIRTs was highlighted as more time and resources are needed to extract relevant information for incident responses~\cite{Jaatun2020}. Problems in having quality data and software (tools) for CSIRTs' operations were highlighted~\cite{Bourgue2013}, calling for more research focusing on tools and data, e.g. tools and data evaluations. Other challenges include insufficient security experts and qualified personnel in CSIRTs~\cite{Krstic2019, Kashiwazaki2018, Oguchi2018, ruefle2014computer}, lack of logging security events and network monitoring~\cite{Jaatun2020} and lack of a systematic preventive measure against cyber threats~\cite{Metzger2011}. Some challenges related to developing tools for incident responses have been addressed~\cite{Kijewski2011, Mokaddem2019, Reyes2018} while many other challenges remain unsolved, hence calling for further research on this topic. 

It is also observed that a lack of a systematic approach to evaluating data and tools at national CSIRTs~\cite{Kassim2022}. This is particularly a problem for free tools and public data, which often do not go through a proper quality assurance process like commercial tools and closed-source data. It is often the case that tools and data are checked on an ad-hoc and informal basis, e.g., by checking with peers, via an Internet search, or by conducting some lightweight self-testing~\cite{Kassim2022}. Hence, a systematic evaluation of tools and data is a crucial area in the operations that needs attention~\cite{Kassim2022}, echoing what was reported in~\cite{Bourgue2013}. This helps to address the lack of effective national computer-related incident response organisations in some nations to address the overall cyber domain ~\cite{hossain2021cyber}.

\section{Discussions}
\label{sec:Discussions-SLR}

This section discusses key findings from the SLR, research gaps, and recommendations for future research. 

\subsection{Key Findings}
The key findings and insights from the SLR can be summarised into several points; 1) the use of public data and free tools are ubiquitous within national CSIRTs, probably because of the free and open-source nature of many relevant software such as Linux distributions and digital forensics tools; 2) most research and studies in the literature concerning the use of public data and free tools in national CSIRTs are too generic, incomplete, ad hoc, or fragmented, \emph{which answers RQ1 regarding how has past research studied the current practices in national CSIRTs concerning public data and free tools in national CSIRTs}; 3) national CSIRTs are very active in developing and using free tools for enhancing incident response and information sharing across-national CSIRTs, represented by some significant software projects - MISP with ongoing development by CIRCL, TARANIS developed by NCSC-NL and IntelMQ developed by CERT.at. This is consistent with the encouragement for CSIRTs to seek out tools that already have been developed within the CSIRT community to prevent duplication and to standardise communication~\cite{IGF2014CSIRT},
4) the need to formulate complete systems and systematic procedures to ensure that operations can run well in national CSIRTs is advocated for national CSIRTs,
5) the mixed use of public and closed-source data is common in national CSIRTs as we found even commercial and free tools consume such data for cyber threat investigation purposes; 6) the mixed use of free and closed-source tools is common in national CSIRTs, and many free tools are recommended for daily operations of national CSIRTs, consistent with the recommendation that national CSIRTs need to prioritise the use of open technology that is free, independent and cost-saving~\cite{haidar2021analysis}; 7) the continuous usage and advocating of free tools and public data in national CSIRTs implies positiveness towards free tools and public data by staff of national CSIRTs \emph{which indirectly answers RQ2 concerning how has past research studied national CSIRT staff's perception on the usefulness of public data and free tools in their daily practice}; 8) more useful information about the use of public data and free tools in the operations of national CSIRTs can be gathered from websites of national CSIRTs and cross-CSIRT organisations than from research papers indicating a gap in current literature; 9) a small number of national CSIRTs (e.g., CIRCL, CERT.at, CERT-EU, JpCERT/CC) have been playing very active role in publicly promoting the use of public data and free tools; 10) the use of public data and free tools are concentrated more on a number of areas such as information sharing across-national CSIRTs , general cyber threat intelligences, digital forensics and malware analysis; 11) a number of public data sources and security feeds play an essential role in the operations of national CSIRTs, including CVE, NVD, malware information, attack sensors, darknet, OSN such as Twitter; 12) a number of national CSIRTs (e.g., CIRCL, JPCERT/CC) also provide public-facing open data and free tools, and many national CSIRTs offer recommendations and advice to organisations and citizens on end user-facing free tools; 13) clearly, there is a general observation that most information on national CSIRTs’ websites is public-facing and not meant to disclose or even discuss what tools are used by national CSIRTs in their operations. There may be some indirect indication that national CSIRTs use the mentioned tools, but no explicit representation because this is not the purpose of the documents published on their websites which justifies why our list of tools identified from the study could be incomplete. 

\subsection{Research Gaps and Recommendations}

The SLR identified several research gaps as highlighted below; 1) lack of research and public reports on staff's understanding and adoption of public data and free tools in national CSIRTs' operations; 2) systematic evidence on the benefit of using free tools and public data to facilitate incident response in national CSIRTs is lacking; 3) lack of case studies demonstrating generalisable procedures involving public data and free tools, to ensure quality of investigation; 4) what most national CSIRTs are doing with public data and free tools in the operations is less known from the research literature; 5) the diversity of free tools and public data seems insufficient, with a lack of understanding of what is needed and what is still missing in national CSIRTs; 6) none of the national CSIRT publications and research papers has empirical evidence of direct coverage on how national CSIRT staff perceive public data and free tools are practically useful, which calls for more empirical research on this topic.

Overall, the key findings and research gaps identified from the study suggest the need for more empirical research concerning tools and data in the operational practices of national CSIRTs. This suggestion is further reinforced by previous studies suggesting more research concerning tools in CSIRTs' operations, including tool development and tool evaluation~\cite{Tondel2014}. Despite CSIRTs generally lacking advanced tools for analysis work~\cite{VanderKleij2017}, this can be further improved through more research activities ~\cite{Jaatun2020} with more scientific rigour. This concerns how technology can best support security incident responses~\cite{Werlinger2010} and subsequently improve national CSIRT practices by formulating procedures for effective operations~\cite{haidar2021analysis}. Therefore, more research is recommended on the topic of national CSIRTs -- a less-studied topic in academic research. This is supported by a previous study that found though there are significant studies on cyber security threats and their countermeasure but very limited study or reference material is available in regards to the CSIRTs and their perspectives within nations~\cite{hossain2021cyber}. 

In a nutshell, the key findings and research gaps identified from the study confirm the need for more empirical research concerning tools and data in the operational practices of national CSIRTs. This includes investigating how public data and free tools can be used more effectively by national CSIRTs and developing standardised and systematic procedures and frameworks for key areas of operational practices at national CSIRTs (e.g., for tool and data evaluation) to ensure the quality of incident responses.

\bibliographystyle{elsarticle-num}
\bibliography{main.bib}

\appendix

\section{National CSIRTs and cross-CSIRT organizations covered in the SLR}
\label{appendix:CSIRTs}

In Phase 1, we identified 125 national CSIRTs' websites from 111 countries representing 5 regions, from which we included 100 national CSIRTs' websites from 86 countries representing 5 regions, listed below (ordered by names, alphabetically). All such websites have either an English section or some English publications.

\begin{enumerate}
\item (Albanian) National Authority for Electronic Certification and Cyber Security (NAECCS / AKCESK, \url{https://cesk.gov.al/publicAnglisht_html/rreth-nesh/index.html})

\item Australian Cyber Security Centre (ACSC, \url{https://www.cyber.gov.au/})

\item National Computer Emergency Response Team of Austria (CERT.at, \url{https://cert.at/})
\item Austrian Government Computer Emergency Response Team (GovCERT Austria, \url{http://www.govcert.gv.at/})

\item CERT Azerbaijan (CERT.AZ, \url{https://cert.az/en/})
\item Azerbaijan Computer Emergency Response Center (CERT.GOV.AZ, \url{https://cert.gov.az/en})

\item Bangladesh Government's e-Government Computer Incident Response Team (BGD e-GOV CIRT, \url{https://www.cirt.gov.bd/})

\item Federal Cyber Emergency Team of Belgium (CERT.be, \url{https://www.cert.be/en})

\item Bhutan Computer Incident Response Team (BtCIRT, \url{https://www.btcirt.bt/})

\item Computer Emergency Response Team Brazil (CERT.br, \url{https://www.cert.br/})
\item (Brazilian) Center for the Treatment of Security Incidents on Computer Networks (CTIR Gov, \url{http://www.ctir.gov.br/})

\item Brunei Computer Emergency Response Team (BruCERT, \url{http://www.brucert.org.bn/})

\item Bulgarian Computer Security Incidents Response Team (CERT Bulgaria, \url{https://www.govcert.bg/EN/Pages/default.aspx})

\item Burkina Faso Computer Incident Response Team (CIRT.BF, \url{https://www.cirt.bf/?lang=en})

\item Cameroon CIRT (\url{https://www.cirt.cm/?language=en})

\item Canadian Centre for Cyber Security (CERT-CA,  \url{https://www.cyber.gc.ca/en/})

\item Caribbean Cyber Emergency Response Team (CARICERT, \url{https://www.caricert.cw/})

\item National Computer Network Emergency Response Technical Team/Coordination Center of China (CNCERT/CC, \url{https://www.cert.org.cn/publish/english/})

\item Croatian National CERT (CERT.hr, \url{https://www.cert.hr/en/home-page/})

\item National CSIRT of Cyprus (CSIRT-CY, \url{https://csirt.cy/en/})

\item Computer Security Incident Response Team of the Czech Republic (CSIRT.CZ, \url{https://nukib.cz/en/cyber-security/})

\item (Danish) Centre for Cyber Security (CFCS, \url{https://cfcs.dk/en/})

\item Egyptian Computer Emergency Readiness Team (EG-CERT, \url{https://www.egcert.eg/})

\item Computer Emergency Response Team Estonia (CERT-EE, \url{https://www.ria.ee/en/cyber-security/cert-ee.html})

\item Ethiopian Cyber Emergency Readiness and Response Team (Ethio-CER$^2$T, \url{http://ethiocert.insa.gov.et/})

\item National Cyber Security Centre Finland (NCSC-FI, \url{https://www.kyberturvallisuuskeskus.fi/en})

\item French Government CSIRT (CERT-FR,  \url{http://www.cert.ssi.gouv.fr/})

\item Computer Emergency Response Team-Georgia (CERT-GOV-GE, \url{https://github.com/CERT-GOV-GE}\footnote{The GitHub web page listed its official website as \url{http://cert.gov.ge/}, which unfortunately was broken when we tried to access it for this SLR.})

\item (German) Computer Emergency Response Team (CERT-Bund, \url{https://www.bsi.bund.de/EN/})

\item Ghana National CERT (CERT-GH, \url{http://www.cert-gh.org/})

\item Hong Kong Computer Emergency Response Coordination Centre (HKCERT, \url{http://www.hkcert.org/})
\item Government CERT Hong Kong (GovCERT.HK, \url{http://www.govcert.gov.hk/})

\item Computer Incident Response Team Iceland (CERT-IS, \url{https://www.cert.is/default.aspx?pageid=6b347de1-b177-11ea-945f-005056bc2afe})

\item Indian Computer Emergency Response Team (CERT-IN, \url{http://www.cert-in.org.in/})

\item Indonesia Computer Emergency Response Team (ID-CERT, \url{https://www.cert.or.id/beranda/en/})
\item Indonesia Security Incident Response Team on Internet Infrastructure/Coordination Center (ID-SIRTII/CC, \url{https://idsirtii.or.id/})

\item Iran Computer Emergency Response Team/ Coordination Center (CERTCC MAHER, \url{https://cert.ir/?language_id=1})

\item Computer Security Incident Response Team for Ireland (CSIRT-IE, \url{https://www.ncsc.gov.ie/CSIRT/})

\item Israeli Cyber Emergency Response Team (CERT-IL, \url{https://www.gov.il/en/departments/news/119en})

\item JPCERT Coordination Center (JPCERT/CC, \url{https://www.jpcert.or.jp/english/})
\item National Center of Incident Readiness and Strategy for cyber security (NISC, \url{https://www.nisc.go.jp/eng/index.html})

\item Kazakhstan CERT (KZ-CERT, \url{https://cert.gov.kz/})

\item Kenya Computer Incident Response Team Coordination Centre (KE-CIRT/CC, \url{https://www.ke-cirt.go.ke/})

\item CERT Coordination Center of Korea (KrCERT/CC, \url{https://www.krcert.or.kr/krcert/intro.do})
\item Korea National Computer Emergency Response Team (KN-CERT, \url{https://eng.nis.go.kr/EAF/1_7.do})

\item Kosovo National CERT (KOS-CERT, \url{https://www.kos-cert.org/en/index.php/home})

\item Kuwait National Cyber Security Center (NCSC-KW, \url{https://citra.gov.kw/sites/en/Pages/cyber security.aspx})

\item Lao Computer Emergency Response Team (LaoCERT, \url{https://www.laocert.gov.la/})

\item Information Technologies Security Incidents Response Institution Latvia (CERT.LV, \url{https://www.cert.lv/en})

\item National Information Security \& Safety Authority (NISSA / LibyaCERT, \url{https://nissa.gov.ly/en/}

\item (Lithuania) National Cyber Security Centre (NKSC or NCSC / formerly CERT-LT, \url{https://www.nksc.lt/en/})

\item Computer Incident Response Center Luxembourg (CIRCL, \url{http://www.circl.lu/})
\item CERT Gouvernemental Luxembourg (GOVCERT.LU,  \url{https://www.govcert.lu/en/})

\item Macau Computer Emergency Response Team - Coordination Centre (MOCERT, \url{https://www.mocert.org/})

\item National Center for Computer Incident Response of the Republic of Macedonia (MKD-CIRT, \url{https://mkd-cirt.mk/?lang=en})

\item Malaysian Computer Emergency Response Team (MyCERT, \url{http://www.mycert.org.my/})

\item Malta National CSIRT (CSIRTMalta, \url{https://maltacip.gov.mt/en/CIP_Structure/Pages/CSIRTMalta.aspx})

\item Computer Emergency Response Team of Mauritius (CERT-MU, \url{http://cert-mu.govmu.org/})

\item (Moldovan) Center for Response on cyber security Incidents (CERT-GOV-MD, \url{https://stisc-cert.gov.md/?lang=en})

\item Mongolia Cyber Emergency Response Team/Coordination Center (MNCERT/CC, \url{https://mncert.org/#/en/})\footnote{On the APCCERT website, the Mongolian CSIRT has a different name -- Mongolian Cyber Incident Response Team (MonCIRT), but its website listed on the APCERT website (\url{http://www.moncirt.org.mn/}) was broken, so we decided to use MNCERT/CC only.}

\item National Montenegrin Computer Incident Response Team (CIRT.ME, \url{http://www.cirt.me/en/cirt?alphabet=lat})

\item Myanmar Computer Emergency Response Team (mmCERT, \url{https://www.mmcert.org.mm/})

\item National Cyber Security Center - Netherlands (NCSC-NL, \url{https://english.ncsc.nl/})

\item CERT New Zealand (CERT NZ, \url{https://www.cert.govt.nz/})
\item New Zealand National Cyber Security Centre (NCSC, \url{https://www.ncsc.govt.nz/})

\item Nigeria Computer Emergency Response Team (ngCERT, \url{https://www.cert.gov.ng/})

\item Norwegian Computer Emergency Response Team (NorCERT, \url{https://nsm.no/areas-of-expertise/cyber-security/norwegian-national-cyber-security-centre-ncsc/})

\item Oman National CERT (OCERT, \url{https://cert.gov.om/})

\item Philippine National Computer Emergency Response Team (CERT-PH, \url{http://ncert.gov.ph/})

\item CERT Polska (CERT.PL, \url{https://www.cert.pl/en/})
\item Republic of Poland Governmental Computer Security Incident Response Team (CSIRT GOV PL, \url{https://csirt.gov.pl/cee})

\item CERT Portugal (CERT.PT, \url{https://www.cncs.gov.pt/en/certpt_en/})

\item Qatar National Center for Information Security (Q-CERT, \url{https://www.qcert.org/})

\item Rwanda National Computer Security and Incident Response Team (RW-CSIRT, \url{http://www.rw-csirt.rw/eng/})

\item Cyber Security and Incident Response Team for the governmental networks of the Russian Federation (CERT.GOV.RU, \url{http://www.gov-cert.ru/en/})

\item Saudi CERT (\url{https://nca.gov.sa/en/pages/cert.html})

\item National CERT of the Republic of Serbia (SRB-CERT, \url{https://www.cert.rs/en})

\item Singapore Computer Emergency Response Team (SingCERT, \url{https://www.csa.gov.sg/singcert}

\item National Cyber Security Centre SK-CERT (formerly known as Slovak Computer Emergency Response Team, \url{https://www.sk-cert.sk/en/})
\item (Slovakian) National Agency for Network and Electronic Services (GOV CERT SK, \url{https://www.cert.gov.sk/en/})

\item Slovenian National Cyber Security Incident Response Centre (SI-CERT, \url{https://www.cert.si/en/})

\item (South African) National cyber security Hub / (South African National CSIRT, \url{https://www.cyber securityhub.gov.za/})

\item Spanish cyber security and Incident Management Teams (CSIRT.es, \url{https://www.csirt.es/index.php/en/})
\item National cyber security Institute of Spain (INCIBE-CERT, \url{https://www.incibe-cert.es/en})

\item Sri Lanka CERT|CC (\url{https://www.cert.gov.lk/})

\item Sudan Computer Emergency Response Team (Sudan CERT, \url{http://www.cert.sd/})

\item National Cyber Security Centre Switzerland (NCSC.ch, \url{https://www.ncsc.admin.ch/melani/en/home.html})

\item National CSIRT of Switzerland (SWITCH Computer Emergency Response Team, \url{https://www.switch.ch/})
\item Taiwan National Computer Emergency Response Team / National Center for Cyber Security Technology (TWNCERT / NCCST, \url{https://www.twncert.org.tw/})
\item Taiwan Computer Emergency Response Team/Coordination Center (TWCERT/CC, \url{https://www.twcert.org.tw/en/mp-2.html})

\item Tanzania Computer Emergence Response Team (TZ-CERT, \url{https://www.tzcert.go.tz/})

\item Thailand Computer Emergency Response Team (ThaiCERT, \url{https://www.thaicert.or.th/about-en.html})

\item Tonga National Computer Emergency Response Team (CERT Tonga, \url{https://www.cert.gov.to/})

\item (Turkish) National Computer Emergency Response Center (USOM / TR-CERT, \url{https://www.usom.gov.tr/})

\item UAE Computer Emergency Response Team (aeCERT, \url{https://www.tra.gov.ae/aecert/en/})

\item Uganda Computer Emergency Response Team (Ug-CERT, \url{https://www.ug-cert.ug/})

\item Computer Emergency Response Team of Ukraine (CERT-UA, \url{https://cert.gov.ua/})

\item (UK) National Cyber Security Centre (NCSC UK, \url{https://www.ncsc.gov.uk/})

\item (US) Department of Homeland Security cyber security and Infrastructure Security Agency (CISA / US-CERT, \url{https://us-cert.cisa.gov/})

\item Zambia Computer Incident Response Team (ZM CIRT, \url{http://www.cirt.zm/})
\end{enumerate}

In addition, we also identified the following 11 cross-CSIRT organizations, which are all included in our SLR:
\begin{enumerate}
\item CERT-EU (\url{https://cert.europa.eu/})
\item European Union Agency for cyber security (ENISA, \url{https://www.enisa.europa.eu/})

\item CMU CERT/CC (\url{https://www.sei.cmu.edu/about/divisions/cert/})

\item Forum of Incident Response Team (FIRST, \url{https://first.org/})

\item Asia Pacific CERT (APCERT, \url{http://www.apcert.org/})

\item Organisation of The Islamic Cooperation – Computer Emergency Response Teams (OIC-CERT, \url{https://www.oic-cert.org/})

\item African Computer Security Incident Response Teams (AfricaCERT, \url{https://www.africacert.org/})

\item European Government CERTs (EGC) group (\url{http://www.egc-group.org/})

\item Task Force on Computer Security Incident Response Teams (TF-CSIRT, \url{https://tf-csirt.org/})

\item Trusted Introducer Service (\url{https://www.trusted-introducer.org/})

\item International Telecommunication Union (ITU), (\url{https://www.itu.int/})
\end{enumerate}

In the following, we also list the $125-100=25$ national CSIRTs (each from a different country or region) we identified but excluded because we could not find an English section on their official websites or their websites included only a few non-technical web pages in English.
\begin{enumerate}
\item CSIRT of Argentine Ministry of Security (MINSEG-CSIRT, \url{https://csirt.minseg.gob.ar/})

\item Computer Security Incident Response Team of the Republic of Benin (bjCSIRT, \url{https://csirt.gouv.bj/})

\item Bosnia and Herzegovina Computer Emergency Response Team (BIH CERT)

\item National Cambodia Computer Emergency Response Team (CamCERT, \url{https://www.camcert.gov.kh/en/category/activities/trainingseminar/})

\item Chilean Computer Emergency Response Team (CLCERT, \url{https://www.clcert.cl/})

\item Grupo de Respuesta a Emergencias Cibernéticas de Colombia (colCERT, \url{http://www.colcert.gov.co/})

\item Cote d'Ivoire Computer Emergency Response Team (CI-CERT, \url{https://www.cicert.ci/})

\item Centro de Respuesta a Incidentes Informaticos de la Agencia de Regulaciony Control de las Telecomunicaciones de Ecuador (EcuCERT, \url{https://www.ecucert.gob.ec/})

\item Guatemala CSIRT (CSIRT GT, \url{https://twitter.com/csirtgt} and \url{http://csirt.org.gt/})\footnote{CSIRT GT's Twitter account was not actually active and its website was broken at the time of our SLR.}

\item (Hungarian) National Cyber Security Center (NCSC Hungary, \url{https://nki.gov.hu/})

\item Computer Security Incident Response Team - Italia (CSIRT-ITA, \url{https://csirt.gov.it/})

\item Computer Emergency Response Team of Kyrgyz Republic (CERT-KG, \url{http://cert.gov.kg/})

\item Computer Emergency Response Team Mexico (CERT-MX, \url{http://www.cns.gob.mx/})

\item Monaco Government CERT (CERT-MC, \url{https://amsn.gouv.mc/})

\item Moroccan Computer Emergency Response Team (maCERT, \url{https://www.dgssi.gov.ma/fr/macert.html})

\item Norwegian Communications Authority (EkomCERT, \url{https://www.nkom.no/english})

\item Computer Security Incident Response Team Panama (CSIRT Panama, \url{https://cert.pa/})

\item Paraguay Equipo de Respuesta ante Incidentes Ciberneticos (CERT-PY, \url{https://www.cert.gov.py/})

\item Peru CERT (PeCERT, \url{https://www.pecert.gob.pe/})

\item Romanian National Computer Security Incident Response Team (CERT-RO, \url{https://www.cert.ro/})

\item (Swedish) CERT-SE (\url{https://www.cert.se/})

\item Tunisian Computer Emergency Response Team (tunCERT, \url{https://www.ansi.tn/node/3388})

\item National Computer Security Incident Response Center Uruguay (CERTuy, \url{http://www.cert.uy/})

\item Uzbekistan Computer Emergency Response Team (UzCERT, \url{https://uzcert.uz/})

\item Vietnam Computer Emergency Response Team (VNCERT, \url{http://www.vncert.gov.vn/})
\end{enumerate}

\end{document}